\documentclass[final]{agujournal2019}
\usepackage{url} 
\usepackage{soul}
\usepackage{amsmath,amssymb}
\draftfalse

\journalname{Geophysical Research Letters}

\newcommand{\ie}{i.e.\,} 
\newcommand{\eg}{e.g.\,} 
\newcommand{\Pran}{\mbox{\textit{Pr}}} 
\newcommand{\Ek}{\mbox{\textit{Ek}}}
\newcommand{\Ra}{\mbox{\textit{Ra}}}
\newcommand{\Ro}{\mbox{\textit{Ro}}}

\newcommand{\Rm}{\mbox{\textit{Rm}}}

\newcommand{\Lc}{\mathcal{L}}
\newcommand{\U}{\mathcal{U}}
\newcommand{\lu}{\ell}
\newcommand{\lnu}{\ell_{\nu}}
\newcommand{\lt}{\ell_{i}}
\newcommand{\vel}{\boldsymbol{u}}
\newcommand{\vect}[1]{\boldsymbol{#1}}
\newcommand{\vorz}{\zeta}
\newcommand{\vor}{\boldsymbol{\omega}}
\newcommand{\velg}{\boldsymbol{u}_{\perp}}
\newcommand{\moyz}[1]{\left \langle #1 \right \rangle}

\begin{document}

%
%


\title{The cross-over from viscous to inertial lengthscales in rapidly-rotating convection}

%
%




\authors{C. Guervilly\affil{1}, and E. Dormy\affil{2}}
\affiliation{1}{School of Mathematics, Statistics and Physics, Newcastle University, Newcastle Upon Tyne NE1 7RU, United Kingdom}
\affiliation{2}{Department of Mathematics \& its Applications, \'Ecole Normale Sup\'erieure, CNRS, PSL University, 75005 Paris, France}




\correspondingauthor{C\'eline Guervilly}{celine.guervilly@newcastle.ac.uk{ ; \rm Emmanuel Dormy,} {emmanuel.dormy@ens.fr}}




\begin{keypoints}
\item The dominant flow lengthscale of rapidly-rotating convection follows either the viscous or the inertial scales depending on the flow speed. 
\item The cross-over between the two regimes occurs when the Rossby number measuring the flow speed exceeds the Ekman number to the power 2/3.
\item The Rossby number in the Earth's core is much greater than this threshhold value.
\end{keypoints}

%
%

%
%

\begin{abstract}
Convection is the main heat transport mechanism in the Earth’s liquid core and is thought to power the dynamo that generates the geomagnetic field. Core convection is strongly constrained by rotation while being turbulent. Given the difficulty in modelling these conditions, some key properties of core convection are still debated, including the dominant energy-carrying lengthscale. Different regimes of rapidly-rotating, unmagnetised, turbulent convection exist depending on the importance of viscous and inertial forces in the dynamics, and hence different theoretical predictions for the dominant flow lengthscale have been proposed. Here we study the transition from viscously-dominated to inertia-dominated regimes using numerical simulations in spherical and planar geometries. We find that the cross-over occurs when the inertial lengthscale approximately equals the viscous lengthscale. This suggests that core convection in the absence of magnetic fields is dominated by the inertial scale, which is hundred times larger than the viscous scale.

\end{abstract}

\section*{Plain Language Summary}
Convection occurs in the Earth’s core due to local changes of the fluid density and is a key process for the evolution and habitability of our planet. Indeed, the convective motions of the electrically conducting fluid generates the geomagnetic field and contributes to the thermal and chemical mixing in the core. Since we have relatively few observations of the deep Earth interior, the properties of core convection are still not fully understood.  Convective motions are strongly constrained by the rotation of the planet and turbulent, which makes realistic core conditions difficult to model. Here, we study one important property of convection: the lengthscale at which the convective flows are the most energetic. Rotation constrains the flows to develop into columnar flows that are aligned with the rotation axis. The transverse lengthscale of these columns can vary considerably depending on the strength of the viscous and inertial forces notably. Using numerical models, we show that the dominant convective lengthscale follows distinct theoretical scaling depending on the flow speed. This allows us to predict that, in the absence of magnetic fields, the Earth’s core is in the inertial regime, where the dominant convective lengthscale is of the order of $10$km. 

%
%

%


%
%
%
%

\section{Introduction}

The Earth's magnetic field originates from the liquid outer core, where a hydromagnetic dynamo converts the kinetic energy provided by convection into magnetic energy. 
The convective origin of the geodynamo is widely accepted, 
but we have relatively few
observations revealing the dynamics within the liquid core so the details of 
core convection and the dynamo process are still actively debated \cite{Landeau2022}. 
An important issue concerns the role played by the magnetic field in shaping the 
convective flows (\eg \citeA{Dormy2016,Yadav1206b,Hughes2016,Aubert2017}).
In particular, it has been suggested recently that the presence of a strong initial magnetic field (imposed externally)
might be necessary to kickstart the geodynamo in the early Earth history \cite{Cattaneo2022}.
In other words, is unmagnetized convection able to generate a magnetic field from a seed field of small amplitude in core conditions? 

To understand this point, we need to consider the different dynamical regimes in which 
convection might be operating in core conditions in the absence of magnetic fields.
Dynamical regimes are usually defined by the dominant
force balances in the Navier-Stokes equation, which governs the fluid motions. 
For rapidly-rotating flows, the primary force
balance in the fluid interior
is expected to be the geostrophic balance between the Coriolis force (produced by 
the rotation of the reference frame) and the pressure force. The geostrophic balance
is thought to be relevant for the Earth's core at large scales
because the rotation timescale ($1/\Omega\sim1$d, where $\Omega$ is the planetary rotation
rate) is much shorter than both the turnover timescale of convection 
($\Lc/\U\sim100$yr, where $\Lc$ is the core scale and $\U$ a typical flow speed) 
and the viscous timescale ($\Lc^2/\nu\sim100$Gyr, where $\nu$ is the kinematic viscosity), 
so inertia and viscous forces play a secondary role in the dynamics.
The rotational constraint can be quantified by a Rossby number 
\mbox{$\Ro=\U/\Omega\Lc\ll1$} and an Ekman number \mbox{$\Ek=\nu/\Omega \Lc^2\ll1$}. 
Under the conditions \mbox{$\Ro\ll1$} 
and \mbox{$\Ek\ll1$},
the flow is organised
into columns, becoming nearly invariant along the direction of the rotation axis.
Rapidly-rotating convection requires a deviation from the geostrophic balance 
and is governed by a secondary force balance in the Navier-Stokes equation, which involves 
the buoyancy force produced by density perturbations. 
Two distinct secondary force balances have been suggested for rapidly-rotating unmagnetized convection: 
the viscous balance, which involves the viscous force, the buoyancy force and the 
Coriolis force, and the inertial balance, where the nonlinear inertia
enters the secondary force balance instead of the viscous force \cite{Jones2015TOG}. 
One of the most notable differences between these two dynamical regimes is the dominant transverse (\ie non-axial) lengthscale of the convective columnar flows.
Here by ``dominant" scale, we mean the most energetic scale of the
flow velocity $\vel$.
The lengthscale associated with the different regimes can be directly estimated from scaling arguments in the vorticity equation, where the vorticity $\boldsymbol{\omega}=\nabla\times\vel$ is a local measure of the fluid rotation (\eg \ \citeA{Aurnou2020}). 
In the viscous regime, the balance between the Coriolis and viscous terms is achieved at the viscous lengthscale $\lnu$:
\begin{equation} 
   2 \Omega \frac{\partial \vel}{\partial z} \sim \nu \nabla^2 \boldsymbol{\omega}~~\Rightarrow ~~ 
   \frac{2\Omega \U}{H} \sim \frac{\nu \U}{\lnu^3} ~~\Rightarrow ~~
    \frac{\lnu}{H} \sim \Ek^{1/3},
    \label{eq:lnu}
\end{equation}
where we assume that the axial gradients of $\vel$ scale with the height of the column ($H\sim\Lc$), while the 
transverse gradients scale with $\lnu$.
In the inertial regime, the balance between the Coriolis and inertial terms is achieved at the inertial lengthscale $\lt$:
\begin{equation}
    2 \Omega \frac{\partial \vel}{\partial z} \sim \vel \cdot \nabla \boldsymbol{\omega} ~~\Rightarrow ~~
    \frac{2\Omega \U}{H} \sim \frac{\U^2}{\lt^2} ~~\Rightarrow ~~ 
    \frac{\lt}{H} \sim \Ro^{1/2}.
    \label{eq:lt}
\end{equation}
By contrast with the viscous lengthscale, $\lt$ therefore increases with 
the flow speed.
In the Earth's core where $\Ek\approx10^{-15}$ and $\Ro\approx10^{-6}$, 
these lengthscales differ by two orders of magnitude with $\lnu \simeq 10^{-5} \Lc \simeq 30 \, {\rm m}$, while
$\lt \simeq 10^{-3} \Lc \simeq 3 \, {\rm km}$. 
This difference is very significant for the geodynamo when considering one basic requirement of dynamo action encapsulated by the magnetic 
Reynolds number, which compares magnetic diffusion and induction timescales. 
In mean-field dynamo theory, if the magnetic field grows at the large scale $\Lc$ driven by a flow at the small scale $\ell$, the relevant magnetic Reynolds number corresponds to the
geometrical mean of both lenghtscales $\Rm_{\rm mf} = \U \sqrt{\Lc\lu}/\eta$,
where $\eta$ is the magnetic diffusion \cite{Mof19}.
Values of the magnetic Reynolds number of at least $10$ are often considered necessary 
for dynamo action (\eg \ \citeA{Backus1958,Luo2020}). 
Using a typical estimate of $\Rm_\Lc = \U \Lc/\eta \approx 10^3$ in the Earth's core implies that $\Rm_{\rm mf}$ 
remains greater than $10$ down to scales as small as $\ell\sim 10^{-4} \Lc$. 
Whilst mean-field dynamos operating at smaller scales have been proposed, they are often built using assumptions
that limit their applicability to core conditions \cite{Childress1972,Calkins2015,Yan2022}.
In this context, the presence of a strong external magnetic field is helpful to get dynamo action started because the modification of the force balance due to Lorentz forces leads to magnetized convection emerging on much larger scales (\eg \citeA{Cha61,Elt72,Stellmach04,Mason2022}).
However, this external field is not necessarily needed to get the geodynamo started if
convection is in the inertial regime because
the theoretical inertial scale $\lt$ is ten times larger than the cut-off scale for mean-field dynamo action.
This point illustrates that finding the relevant dominant lengthscale of 
unmagnetized rapidly-rotating convection is essential 
to understand the generation of magnetic fields in the Earth and in other planets.

Rapidly-rotating convection has been extensively studied using numerical models
and laboratory experiments (\eg \citeA{Cheng2015,Aurnou2020,Kunnen2021,Gastine2023}) and the evolution of the dominant flow lengthscale 
with the parameters is often used as an indication of the dynamical regime. 
The measured lengthscales have been found to follow power laws close to the viscous
scaling \cite{Oliver2023}, the inertial scaling \cite{Aub01,Barker2014,Guervilly2019,Hadjerci2024,Song2024},
or to have a weaker dependence on the Rossby number \cite{Gastine2016,Long2020,Madonia2021,Nicoski2024}. 
The conditions under which a given scaling is preferred remain unclear:
for instance, in planar layer simulations at low $\Ro$, \citeA{Oliver2023} recently found that the flow 
lengthscale remains controlled by viscosity even for strong buoyancy forcings, 
in contrast to the 
simulations of \citeA{Hadjerci2024,Song2024} under similar conditions;
in rotating convection experiments, \citeA{Abbate2023} found that the inertial lengthscale 
as estimated from velocity measurements remains close to the viscous lengthscale
and argued that achieving a significant separation between the two scales is unlikely for the parameter range 
accessible in present day studies.
An important body of work on low Rossby number convection has been built on reduced models  
that use the viscous scale as the transverse length of the convection columns \cite{Sprague2006,Julien2012}.
Therefore, in addition to its importance for dynamo action,
the question of which convection regime and lengthscale are preferred in the core is crucial
for the modelling of the fluid dynamics of planetary cores, as 
well as for chemical mixing, heat transport efficiency and core energy budgets.
In this Letter, we study the behaviour of the dominant flow lengthscales in
numerical simulations of rapidly-rotating convection to determine the conditions under which
they follow theoretical predictions. 
We test whether these conditions depend on the geometry of the system by using both spherical and planar geometries and on 
the fluid properties as measured by the Prandtl 
number (the ratio of the diffusion coefficients as defined below). 
We propose a new criterion to predict which lengthscale is preferred based on the values of the Ekman and Rossby numbers. 

\section{Methods}

\subsection{Planar model}
\label{sec:planar}
In the planar model, convection is driven by imposing a temperature difference $\Delta T$ between 
the bottom and top boundaries, which are distant by a height $d$. 
The gravitational acceleration is $\vect{g} = - g \vect{e}_z$,
and the rotation vector is $\Omega \vect{e}_z$, where $g$ and $\Omega$ are constant and $\vect{e}_z$ is directed upwards in Cartesian coordinates $(x, y, z)$.
The boundary conditions are no-slip and isothermal at the top and bottom of the domain 
and periodic in the horizontal directions. We assume a Boussinesq fluid, with 
kinematic viscosity $\nu$, thermal diffusivity $\kappa$ and 
thermal expansion coefficient $\alpha$, all of which are constant. We solve the Navier-Stokes and temperature 
equations in dimensionless form, obtained by scaling lengths with $d$, times with $d^2/\nu$, 
and temperature with $\Pran \Delta T$. 
The system of dimensionless governing equations is
\begin{eqnarray}
        &&  \frac{\partial \vel}{\partial t} + (\vel \cdot \nabla) \vel + \frac{2}{\Ek} \vect{e}_z \times \vel = -\nabla p + \Ra \theta \vect{\gamma} + \nabla^2 \vel \,
	\label{eq:u}
	\\
	&& \nabla \cdot \vel = 0 ,
	\\
	&& \frac{\partial \theta}{\partial t} + \vel \cdot  \nabla \theta + \vel \cdot \vect{\beta} = \frac{1}{\Pran} \nabla^2 \theta ,
	\label{eq:theta}
\end{eqnarray}
where $\vel=(u_x,u_y,u_z)$ is the velocity field, $p$ the pressure, and $\theta$ the temperature perturbation relative to a background profile. $\vect{\beta}=-\vect{e}_z$ is the vertical gradient of background temperature and $\vect{\gamma}=\vect{e}_z$. 
The dimensionless parameters are the Rayleigh number, $\Ra = \alpha g \Delta T d^3/(\kappa \nu)$,
the Ekman number, $\Ek = \nu/(\Omega d^2)$, and the Prandtl number, $\Pran = \nu/\kappa$.

We perform 3D numerical simulations with $\Ek\in[2\times 10^{-6}, 2\times 10^{-4}]$ and $\Pran=1$. 
The horizontal box width, $h$, is chosen to be least 10 times the horizontal lengthscale of a convective column at onset
\ie $h=\{4d, d, 0.5d\}$ for $Ek=\{2\times 10^{-4},2\times 10^{-5}, 2\times 10^{-6}\}$ respectively. 
Additionally, we present a number of simulations with $h=2d$ for $Ek=2\times 10^{-5}$ to 
check that the results are not affected by a horizontal box confinement. 
For each $\Ek$, we vary the buoyancy forcing via $\Ra$, which takes values close to the onset of convection up to 
values of the reduced Rayleigh number $\widetilde{\Ra}=\Ra\Ek^{4/3}$ greater than a thousand (see Table S1 of the 
supporting information).
All the simulations were performed with the open-source code Dedalus \cite{Dedalus2020}, using
a Fourier decomposition in the horizontal directions and Chebyshev polynomials in the vertical direction.
The numerical resolution are given in Table S1 of the supporting information.
Note that large-scale vortices growing to the box size are known to form in rotating planar convection \cite{Guervilly2014,Favier2014,Rubio2014},
but these are not present here as we use no-slip boundary conditions and relatively moderate values of the Ekman number \cite{Stellmach2014,AguirreGuzman2020}. 

\subsection{Spherical model}
To study rapidly-rotating convection in spherical geometry, we use part of the data sets published in \citeA{Guervilly2019}, 
which consists of 3D simulations ($\Ek\in[10^{-8}, 10^{-6}]$) 
and Quasi-Geostrophic (QG) simulations ($\Ek=10^{-8}$)
in a full sphere geometry with homogeneous internal heating. 
These data sets were obtained at low Prandtl numbers, $\Pran=\{0.01,0.1\}$,
which correspond to values relevant for thermal convection in  liquid metals.
One of the main differences with the planar model is that the gravitational acceleration is radial 
and increases linearly with radius $r$, $\vect{g} = - g' r\vect{e}_r$, in spherical polar coordinates $(r, \theta, \phi)$
and where $g'$ is constant.
We use no-slip and isothermal boundary conditions at the outer boundary.
The 3D simulations were performed with the open-source code XSHELLS \cite{Sch13,Kaplan2017},
which solves the system of dimensionless equations~\eqref{eq:u}-\eqref{eq:theta}, with $\vect{\gamma}=\vect{r}$ and 
$\vect{\beta}=-2r/\Pran \vect{e}_r$
(i.e. gravity associated with a fluid of uniform density and a uniform distribution of heat sources).
The dimensionless numbers are defined in the same manner as in \S\ref{sec:planar}, using 
the radius of the outer sphere $r_o$ as unit for lengths 
and substituting $g\Delta T$ with $g'Sr_o^3/(6\rho C_p \kappa)$ for the definition of $\Ra$, where $S$ is the internal volumetric heating, $\rho$ the density, and $C_p$ the heat capacity at constant pressure. 
XSHELLS uses finite differences in the radial direction and spherical harmonic expansion
in the angular directions. The numerical resolutions used in the 
3D simulations are given in \citeA{Guervilly2019}.

Since 3D simulations at low Ekman numbers are computationally costly, the 3D data sets is complemented with QG simulations. This allows us to test how variations of $\Ra$ and $\Pran$ influence the dynamics. 
The QG model is a 2D numerical model that takes advantage of the rotational constraint and 
assumes that the axial vorticity and the temperature are
invariant along the rotation axis at low Rossby numbers. 
The use of this QG approximation to model
spherical convection at low Rossby numbers and moderate convective forcing is well supported by 
comparison with the 3D results \cite{Guervilly2019,Barrois2022}. 
QG convection is driven by the radial component of gravity, \mbox{$-g' s$}, in cylindrical polar coordinates $(s,\phi,z)$.
The QG model solves the equation for $z$-average axial vorticity $\vorz=(\nabla \times \vect{u})\cdot \vect{e}_z$:
\begin{equation}
	\frac{\partial \vorz}{\partial t} + \left( \velg \cdot \nabla_{\perp} \right) \vorz 
	- \left( \frac{2}{\Ek} + \vorz \right) \moyz{\frac{\partial u_z}{\partial z}}
	= \nabla_{\perp}^2 \vorz - \Ra \moyz{\frac{\partial \theta}{\partial \phi}},
	\label{eq:vorz}
\end{equation}
where $\velg=(u_s,u_{\phi},0)$, $\nabla_{\perp} f \equiv (\partial_s f, \partial_{\phi} f/s,0 )$,
$\nabla^2_{\perp} f \equiv \partial^2_s f + s^{-1} \partial_s f + s^{-2} \partial^2_{\phi} f$,
and the angle brackets denote an axial average between $\pm H$ with $H=\sqrt{1-s^2}$ the 
height of the spherical boundary from the equatorial plane.
The axial velocity $u_z$ is assumed to be linear in $z$ and has two contributions: 
the main contribution comes from mass conservation at $z=\pm H$ and is proportional to 
$\beta=H^{-1}{\rm d H}/{\rm d z}$; 
the other contribution accounts for a parameterized Ekman friction due to the viscous boundary layer.
Additionally, the model solves the equation for the $z$-averaged temperature perturbation, 
assuming that $\theta$ is invariant along $z$.
Further details about the formulation of the model and the numerical resolution of the simulations are available in \citeA{Guervilly2019}. 

Numerical simulations of core dynamics often use Prandtl numbers of order unity 
and spherical shells (\ie include an inner core). 
Therefore, in order to relate our results to the literature, 
we complement the existing data sets with new QG simulations at $\Ek=10^{-8}$
and $\Pran=\{0.01,1\}$
in a spherical shell geometry of aspect ratio $r_i/r_o=0.35$, where $r_i$ is 
the radius of the inner sphere.
Convection is driven by an imposed temperature difference between the inner and outer boundaries.
The numerical resolution and output parameters  of the new QG simulations
are given in Table S2 of the supporting information.

\section{Results}

\subsection{Lengthscales}

\begin{figure}
\centering
\includegraphics[width=\textwidth]{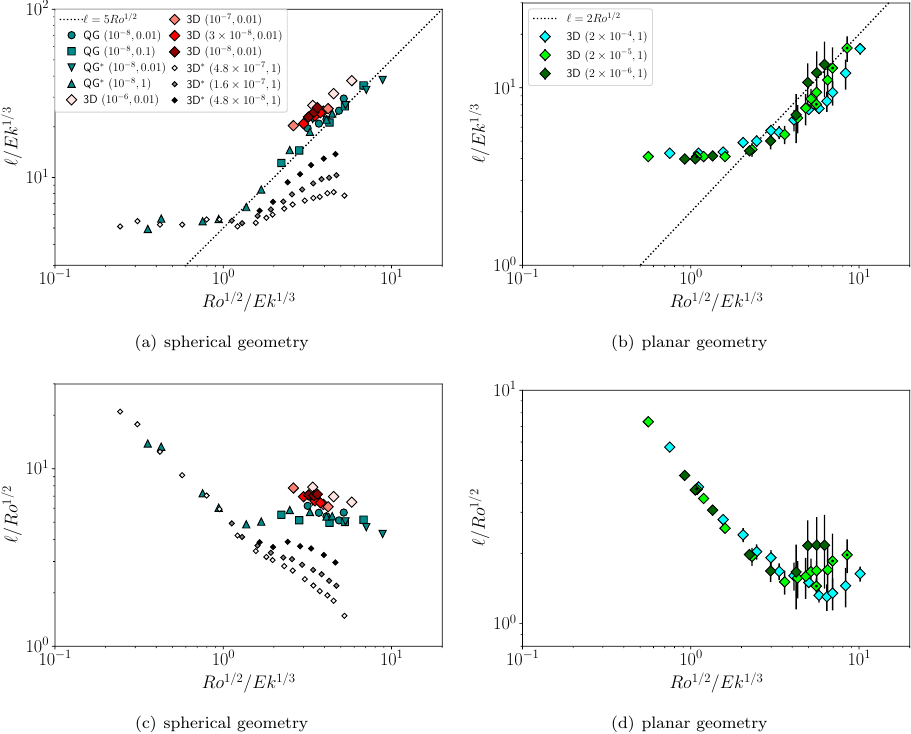}
\caption{
\label{fig:scales}
Dominant flow lengthscale normalised by the viscous lengthscale $\lnu=\Ek^{1/3}$ 
(top row) and by the inertial lengthscale $\lt=\Ro^{1/2}$ (bottom row)
as a function of the ratio $\lt/\lnu=\Ro^{1/2}/\Ek^{1/3}$ in simulations of rotating convection at different $(\Ek, \Pran)$ as indicated in the legend in  spherical geometry (left) and planar geometry (right). 
In (a) and (c), the QG and 3D simulations at $\Pran<1$ are taken from the data sets of \citeA{Guervilly2019} in a full sphere with internal heating.
The QG simulations indicated by an asterisk in the legend entry are performed in a spherical shell (aspect ratio $r_i/r_o=0.35$) with differential heating. 
The 3D simulations at $\Pran=1$ indicated by an asterisk are taken from
\citeA{Gastine2016} in a spherical shell (aspect ratio $r_i/r_o=0.6$) with differential heating.
The Ekman number in \citeA{Gastine2016} has been rescaled to match our definition that uses $r_o$ as unit length.
In (b) and (d), the data points with a central dot indicate simulations ran in a wider box ($h=2d$ for $\Ek=2\times 10^{-5}$).
}
\end{figure}

Figure~\ref{fig:scales} shows the dominant flow lengthscale $\lu$ measured in the
simulations normalised by the viscous scale $\lnu=\Ek^{1/3}$ 
as a function of the ratio $\lt/\lnu$ where we used the inertial scale $\lt=\Ro^{1/2}$.
The Rossby number is an output value from the simulations; for a series with fixed $(\Ek,\Pran)$, 
an increase of $\Ro$ corresponds to an increase in the Rayleigh number.
The Rossby number is based on the root mean square (r.m.s.) value of the radial 
velocity for the spherical simulations and the vertical velocity for the planar simulations. 
The dominant lengthscale $\lu$ corresponds to the
peak of the kinetic energy spectrum.
In the spherical simulations,
$\lu$ is calculated as $\lu=\langle w(r)\pi r/m_p(r) \rangle$, 
where $m_p(r)$ is the azimuthal mode at the peak of time-averaged radial kinetic energy spectra at radius $r$,
$w(r)$ is a weighting factor proportional to the r.m.s. radial velocity at radius $r$ and the angle brackets denote a radial average. 
In spherical simulations for $\lt/\lnu\lesssim1$,
Fig.~\ref{fig:scales}a shows that $\lu$ is essentially unchanged when the 
Rossby number increases, remaining close to the viscous lengthscale $\lnu$. 
For $\lt/\lnu>1$, $\lu$ increases with $\Ro$, following closely the theoretical inertial scaling $\Ro^{1/2}$. 
The fit to the theoretical scaling can be better quantified by normalising $\ell$ by $\Ro^{1/2}$, as shown in Fig.~\ref{fig:scales}c, where the data level on a plateau for $\lt/\lnu>1$.
We therefore observe a transition at $\lt/\lnu\approx 1$:
for $\lt/\lnu<1$, the dominant lengthscale of the convection follows $\lnu$,
whilst for $\lt/\lnu>1$, it follows $\lt$.
The dominant lengthscale for a given simulation therefore corresponds to the larger of the theoretical scales.
For $\Pran=1$, the transition from viscous to inertial regimes is continuous. 
At $\Pran<1$, 
the bifurcation at the onset of convection is subcritical at low Ekman numbers \cite{Guervilly2016,Kaplan2017,Skene2024}:
convection only occurs for large Reynolds numbers and all the solutions are located 
on the inertial branch.
In the inertial regime, the lengthscale is independent of $\Pran$: the series at $\Ek=10^{-8}$ 
superpose well irrespective of the value of $\Pran$, which is varied by two decades. 
In this regime, the lengthscale is also independent of the mode of heating (internal heating in a full sphere 
or differential heating in a spherical shell). 

We compare our data sets in spherical geometry with the results of \citeA{Gastine2016}, which were obtained
in a spherical shell of aspect ratio $r_i/r_o=0.6$ with differential heating and $\Pran=1$.
We select the data sets from \citeauthor{Gastine2016} with the smallest Ekman numbers $\Ek< 10^{-6}$. 
In agreement with our results, the transition from a flow lengthscale independent of $\Ro$ at $\lt<\lnu$
to a dependence of $\Ro$ for $\lt>\lnu$ is clearly visible on the data sets of \citeauthor{Gastine2016},
especially at the smallest Ekman numbers. 
As observed in the spherical simulations of \citeA{Long2020} and \citeA{Nicoski2024},
the slope is less steep than expected from the inertial scaling 
but the data get closer to $\Ro^{1/2}$ at smaller $\Ek$. It is plausible that values of $\Ek\leq 10^{-8}$
for $\Pran=1$ are required to approach the theoretical scaling as seen in the QG simulations.
As shown by the 3D simulations performed at $\Pran=0.01$, the inertial scaling is more easily approached at low $\Pran$. Another consideration to explain the slow convergence of the 3D simulations of \citeA{Gastine2016} towards the inertial scaling is that they used an alternative measurement for the typical flow lengthscale, sometimes called the integral lengthscale,
which is calculated as a weighted average over the whole spectrum \cite{Chr06}.
In our simulations, this integral lengthscale does not capture adequately the dominant
lengthscale of the convection and tends to have a weaker dependence on $\Ro$ (see discussion below).
One last consideration is that we use the spherical harmonics order $m$ to measure the transverse
lengthscale of rotating columnar flows, unlike previous studies, which used the spherical harmonics degree thereby
measuring a lengthscale on a spherical surface. This being said, the cylindrical
radial and azimuthal lengthscales of the convective columns grow with a similar power law in $\Ro$ in the inertial regime \cite{Guervilly2019}, 
so this choice probably does not to affect the results significantly.

In the planar simulations (Fig.~\ref{fig:scales}b),
the dominant flow lengthscale $\lu$ corresponds to the peak of 
the vertical kinetic energy spectrum averaged in $z$
as a function of the horizontal wavenumber $k_h=(k_x^2+k_y^2)^{1/2}$. 
The lengthscale is averaged in time and the standard deviation is indicated by the vertical error bars.
Similarly to the spherical case, $\lu$ remains initially close to the viscous lengthscale until
$\lt/\lnu\approx2$. 
The lengthscale subsequently increases with $\Ro$, following a scaling close to the inertial scaling.
The fit to the theoretical scaling can again be better judged from the plot of
$\lu$ normalised by $\Ro^{1/2}$ (Fig.~\ref{fig:scales}d). The data show a better agreement
with the $\Ro^{1/2}$ scaling at smaller $\Ek$ for $\lt/\lnu \ge 4$.
Datasets at different $\Ek$ do not exactly collapse on each other (with 
$\lu$ varying by less than a factor $1.7$ over two decades of $\Ek$ at fixed $\Ro$), which indicates that a
small dependence of $\lu$ on $\Ek$ remains for these simulations where $\Ek>10^{-6}$. 
Similar results are obtained by \citeA{Song2024} in planar simulations of rotating convection, where 
the dependence of the flow lengthscale on $\Ek$ disappears in the inertial regime for $Ek\lesssim10^{-8}$.
Here we observe the inertial regime over a restricted range of $\Ro$. To widen this
range, we would need to increase the horizontal box size with increasing $\Ra$ to prevent
confinement effects and/or to decrease $\Ek$ to keep low values $\Ro$ as appropriate for core convection.
These constraints make the exploration of the inertial regime increasingly prohibitive in terms of computational resources.

\begin{figure}
\centering
\includegraphics[width=\textwidth]{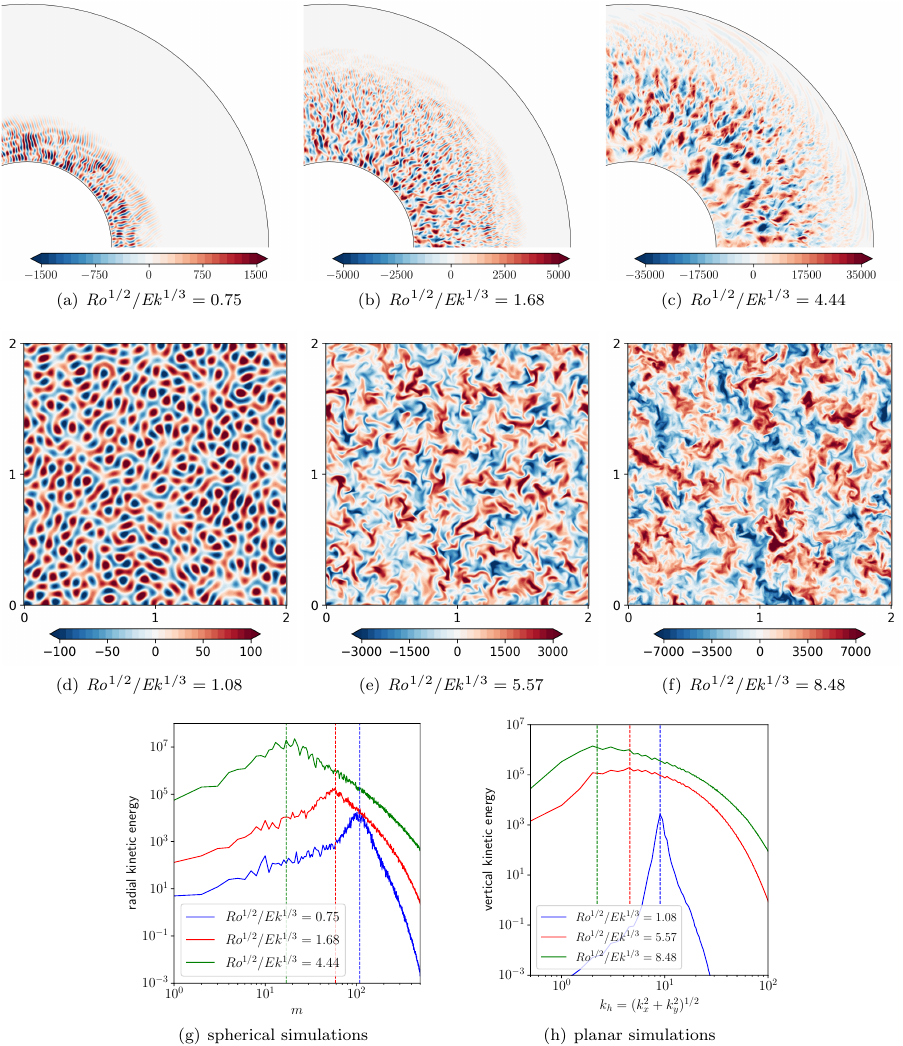}
\caption{(a)-(c) Radial velocity (snapshots) in a quarter of the equatorial plane with increasing Rayleigh numbers from left to right in QG simulations
driven by differential heating in a spherical shell at $(\Ek, \Pran)=(10^{-8}, 1)$;
(d)-(f) vertical velocity (snapshots) in the $xy$ plane at $z=0.25$ 
with increasing Rayleigh numbers from left to right in 3D simulations of planar convection at $(\Ek, \Pran)=(2\times10^{-5}, 1)$ and $h=2d$;
(g) spectra (averaged in time) of the kinetic energy of the radial velocity as a function of the azimuthal mode $m$ at the radius $s=0.4$ in the QG simulations;
(h) spectra (averaged in time and along $z$) of the kinetic energy of the vertical velocity as a function of the horizontal wavenumber $k_h$ in the planar simulations.  
The vertical dashed lines mark the spectral peak.}
\label{fig:u}
\end{figure}

Figure~\ref{fig:u} shows the radial and vertical velocities (respectively) in the equatorial 
and horizontal planes for spherical and planar convection simulations for cases representative of the 
viscous regime (\mbox{$\Ro^{1/2}/\Ek^{1/3}\lesssim1$}, Fig.~\ref{fig:u}(a) and (d)) and the inertial regime (\mbox{$\Ro^{1/2}/\Ek^{1/3}>1$}, Fig.~\ref{fig:u}(b)-(c) and (e)-(f)).
For both geometries, 
the azimuthal and horizontal lengthscales of the convection columns are visibly larger in the inertial regime than in the viscous regime.
This is also seen in the kinetic energy spectra shown in Fig.~\ref{fig:u}(g)-(h),
where the spectral peak visibly shifts to smaller horizontal wavenumbers at larger $\Ro$.
In planar geometry, convection occurs in the whole domain, whereas in spherical geometry, convection is localised 
due the variations in the sloping boundaries. 
In the case of differential heating, convection first develops near the inner sphere, 
where the axial stretching of the columns due to the Coriolis force (which opposes convection) is minimal \cite{Dormy04}. 
As the Rayleigh number increases, the convection occupies a wider region that gradually extends towards the outer sphere. 
In the viscous regime in spherical geometry, the radial flow is, very distinctly, much more extended in the 
cylindrical radial than azimuthal directions.
In the inertial regime, the difference between radial and azimuthal lengthscales is much less pronounced.

Finally, we note that defining a single lengthscale to meaningfully describes the energy distribution 
might not always be straightforward when the kinetic energy spectra is broad, as observed in the inertial regime. 
Here we adopt a definition based on the spectral peak 
because it represents the most energetic lengthscale associated with the velocity. 
In all our simulations, it also corresponds to the lengthscale where  
the convective heat transport takes maximum value, 
as measured on the power spectra of the convective heat flux.
This is illustrated in Figure S1 of the supporting information showing comparison of the power spectra
for both spherical and planar cases in the inertial regime.
The dominant flow lengthscale therefore also dominates the convective heat transport.
Additionally, we measure the horizontal correlation lengthscale based on the auto-correlation function of the vertical velocity in the planar simulations \cite{Nieves2014,Madonia2021}. The comparison
between the correlation lengthscale and the lengthscale based on the spectral peak
shows a good agreement (see Figure S2 in the supporting information).
Alternative measurements for the typical flow lengthscale are also discussed in \citeA{Oliver2023,Hadjerci2024,Song2024a}.
A common choice
is the integral lengthscale that is calculated as a weighted average over the whole spectrum \cite{Chr06}.
In our plane layer simulations (where $\Ek>10^{-6}$), we find that the 
integral lengthscale does not accurately track the behaviour of the dominant energy-carrying lengthscale in the inertial regime, 
as it remains close to values of the viscous lengthscale despite the visible shift of the 
kinetic energy towards lower wavenumbers observed in Fig.~\ref{fig:u}.
The integral scale (using an equivalent definition of \citeA{Chr06} based on
the order $m$) also shows a weaker dependence on $\Ro$ for the spherical QG simulations 
(\eg the best data fit scales as $\Ro^{0.35}$ for $(\Ek,\Pran)=(10^{-8},1)$).
\citeA{Song2024} show that 
the integral scale captures the inertial scaling in planar simulations at very low Ekman numbers ($\Ek \lesssim 10^{-8}$). 
Given that the integral scale incorporates information from the spectrum tail, it is perhaps not surprising 
that it shows a slower convergence towards the inertial scaling than the spectral peak.

To check whether the behaviour of the flow lengthscale is a suitable guide to identify 
the viscous and inertial regimes, we compare the amplitude of the forces in planar and 
spherical simulations in the next section.

\subsection{Force balance}

\begin{figure}
\centering
\includegraphics[width=\textwidth]{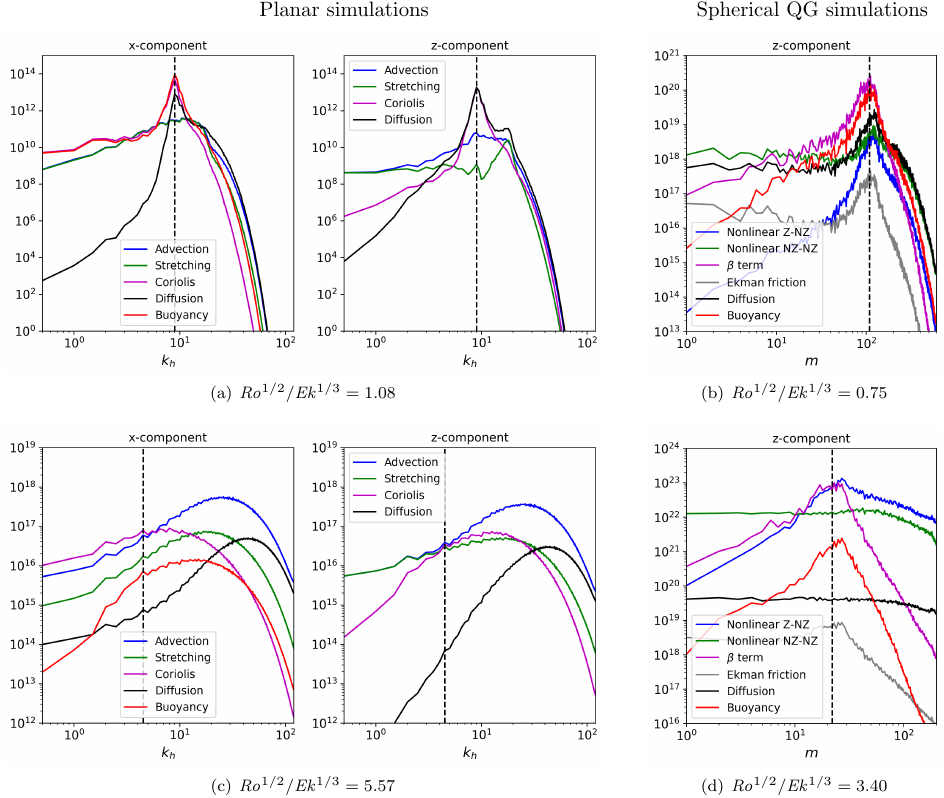}
\caption{Power spectra of the terms in the vorticity equation:
(a) and (c) as a function of the horizontal wavenumber $k_h$ in planar simulations for 
$(\Ek, \Pran)=(2\times10^{-5}, 1)$ and $h=2d$ at $z=0.25$;
(b) and (d) as a function of the azimuthal mode $m$
in spherical QG simulations for $\Ek=10^{-8}$ and (b) $\Pr=1$, (d) $\Pran=0.01$ at $s=0.4$.
The wavenumber corresponding to the spectral peak of the vertical/radial kinetic energy is indicated by a vertical dashed line.}
\label{fig:balance_planar}
\end{figure}

To eliminate the pressure gradient and the gradient part of the forces \cite{Teed2023}, 
we analyse the relative strengths of the terms in the equation for the vorticity, 
$\vor=\nabla \times \vel$. 
We therefore do not study the primary geostrophic balance (which involves a balance between the 
pressure gradient and the Coriolis force), but instead consider the secondary force balance that governs convection.
In planar geometry, the vorticity equation is:
\begin{eqnarray}
    \frac{\partial \vor}{\partial t} + (\vel \cdot \nabla) \vor - (\vor \cdot \nabla) \vel 
    - \frac{2}{\Ek} \frac{\partial \vel}{\partial z} = \Ra \nabla \times (\theta \mathbf{e}_z) + \nabla^2 \vor \, .
	\label{eq:vort}
\end{eqnarray}
Since the force balances are scale-dependent, we look at 
the power spectra of each term in the $x$-, $y$- and $z$-components of equation~\eqref{eq:vort}.
At a given simulated time, we calculate the 2D Fourier transform of each term in the $xy$ plane at depth $z=0.25$.
The squared magnitude of the spectral coefficients is then averaged in time and sorted into 
bins of the horizontal wavenumber $k_h$.
Figure~\ref{fig:balance_planar}(a) and (c) show the spectra for the $x$- and $z$-components of the vorticity as a function of $k_h$ for two planar cases previously identified in the viscous and inertial regimes 
based on the behaviour of the dominant flow lengthscale. 
The $y$-component gives similar result than the $x$-component and is therefore not shown.
The advection and stretching terms correspond to the second and third terms on the left-hand side of equation~\eqref{eq:vort} respectively. 
The case $\Ro^{1/2}/\Ek^{1/3}=1.08$ is representative of the viscous balance: 
the curled Coriolis and buoyancy forces (and to a lesser extent the viscous 
term) have similar amplitude at the spectral peak (corresponding to the lengthscale $\lu$) 
for the $x$-component. For the $z$-component, which is not forced by buoyancy, 
the Coriolis term is balanced by the viscous term, while the non-linear terms 
have much smaller amplitude at lengthscale $\lu$.
For the case $\Ro^{1/2}/\Ek^{1/3}=5.57$, the convection is well above onset 
($\widetilde{\Ra}\approx 200$).
The nonlinear terms are now larger than the viscous term on all wavenumbers 
smaller than $k_h=30$. 
At the energy-carrying lengthscale $\lu$, the Coriolis and nonlinear terms balance  
for the $z$-component and dominates for the $x$-component with a smaller contribution from the buoyancy term, which has a broad spectrum that peaks at scales smaller than the scale $\lu$ (denoted by a vertical dashed line in the figure). 
The viscous term is noticeably smaller than the other terms
at this scale, although its contribution is non-negligible for the $x$-component,
which could explain the small dependence of $\lu$ on $\Ek$ as noted earlier.

We perform a similar analysis for the spherical QG simulations by looking at the power
spectra of each term in the equation for the $z$-averaged axial vorticity (equation~\eqref{eq:vorz}).
Each spectra is computed as a function of the azimuthal mode $m$ (for all $m>0$) at the radius $s=0.4$
and averaged in time. 
Figure~\ref{fig:balance_planar}(b) and (d) show two cases previously
identified in the viscous and inertial regimes. 
The nonlinear term is separated into non-zonal (NZ, \ie $m> 0$) - zonal (Z, \ie $m=0$)  
interactions and NZ-NZ interactions. 
The curled Coriolis force consists of a $\beta$ term due to mass conservation at the boundaries
and a parameterized Ekman friction, which are plotted separately. 
The Ekman friction is a small term for all the non-zonal modes. 
The simulation with $\Ro^{1/2}/Ek^{1/3}\lesssim 1$ gives a similar picture to the planar case, 
where the buoyancy, viscous and $\beta$ terms have similar amplitudes
at the dominant flow lengthscale $\lu$, representing a viscous balance.
The nonlinear terms only have slightly smaller amplitude than the viscous term at that scale. 
The simulation for $\Ro^{1/2}/Ek^{1/3}>1$ is also similar to the planar case, where the $\beta$
and nonlinear terms take comparable values at the lengthscale $\lu$, with a smaller 
contribution from the buoyancy term and even smaller contribution from the viscous term. 

In summary, the spectral representation of the forces in both planar and spherical simulations
shows that, at the dominant flow lengthscale, the force balance for $\Ro^{1/2}/Ek^{1/3}\lesssim 1$ 
is close to the theoretical viscous balance, whilst for $\Ro^{1/2}/Ek^{1/3}> 1$ the balance is more subtle 
than the theoretical inertial balance.  
Indeed the theoretical inertial balance assumes a triple balance for the generation of vorticity, \ie that the Coriolis, inertial and buoyancy terms have similar magnitudes at the dominant flow scale.
In the simulations, we find that the buoyancy term is actually smaller than the 
inertial and Coriolis terms at that scale, suggesting
an upward energy transfer from a smaller injection scale. 
Nevertheless, the dominant flow lengthscale $\lu$ is also the dominant convective lengthscale, corresponding to the spectral peak of the 
convective heat flux as shown in Figure S1 of the supporting information.
The dominance of the Coriolis-inertia balance (secondary to the geostrophic balance)
for the interior dynamics is in agreement
with the results of \citeA{Guzman2021,Oliver2023} using planar simulations of rapidly-rotating convection. 
\citeA{Oliver2023} found that the viscous and buoyancy forces gradually become of
comparable amplitude when the Rayleigh number increases.
This trend is also observed in our simulations at the dominant flow lengthscale. 

Despite the sub-dominance of the buoyancy force in the simulations, 
the behaviour of the dominant flow lengthscale for $\Ro^{1/2}/Ek^{1/3}> 1$ remains consistent with the inertial scaling $\lu\sim\Ro^{1/2}$.
Indeed, this scaling is obtained by considering that the vorticity generation by the Coriolis and the inertial terms are of similar amplitude (equation~\eqref{eq:lt}). 

\section{Conclusions}
Using numerical simulations of rapidly-rotating convection in spherical and planar geometries,
we find that the dominant flow lengthscale increases with the flow speed  
when $\Ro^{1/2}/\Ek^{1/3}>1$ and follows a scaling close to the inertial scaling 
($\lt \sim \Ro^{1/2}$) in this case. 
For $\Ro^{1/2}/\Ek^{1/3}<1$, the flow lengthscale is constant, remaining close to the
viscous lengthscale ($\lnu \sim Ek^{1/3}$).
Therefore, the dominant lengthscale of the convection corresponds 
to the larger of the viscous and inertial lengthscales 
with a cross-over occurring at \mbox{$\lt/\lnu=\Ro^{1/2}/\Ek^{1/3}=\mathcal{O}(1)$}.
Similarly to previous studies of rotating convection \cite{Abbate2023}, our study cannot achieve
a large separation between the viscous and inertial lengthscales, differing by a factor 5 at most
due to computational constraints.
However, the cross-over is within the accessible parameter range and 
we expect future simulations to obtain increasing scale separation as more extreme parameters are modelled.

These results are robust in the sense that they are independent of the
geometry (full sphere, spherical shell, plane layer), mode of heating (internal heating, differential
heating) and Prandtl numbers (varied here between $0.01$ and $1$). The cross-over is also 
observed for all the Ekman numbers considered here between $10^{-4}$ to $10^{-8}$. 
All our simulations used fixed thermal boundary conditions, which are 
arguably
less geophysically relevant
than fixed flux boundaries. However, the studies of \citeA{Calkins2015a,Kolhey2022} in rotating plane-layer convection
and \citeA{Clarte2021} in spherical shells, show that the bulk flow is largely insensitive to the choice of thermal boundary conditions in rapidly-rotating convection at low Ekman numbers, suggesting that the results presented here would generalise to the case of fixed flux boundary conditions.

In the Earth's core, the ratio $\Ro^{1/2}/\Ek^{1/3}$ is estimated to be approximately $100$,
well above the cross-over value. Our results therefore predict that the dominant lengthscale 
of convection is the inertial scale $\lt$ in unmagnetized core conditions.
This implies that a large-scale dynamo could be generated from a small initial seed field in the early Earth history as
the mean-field magnetic Reynolds number at this scale, $\Rm_{{\rm mf}} \approx 30$, could be sufficient.
A strong initial field originating from an external source (such as the moon formation event) 
is therefore not required to kickstart the geodynamo.

\section{Open Research}
Data sets for this research are available on the Figshare powered Newcastle University research data repository (https://data.ncl.ac.uk) \cite{dataset}.






\acknowledgments
C.G. acknowledges support from the UK Natural Environment Research Council under Grants NE/M017893/1. This research made use of the Rocket High Performance Computing service at Newcastle University and the DiRAC Data Intensive service (CSD3) at the University of Cambridge, managed by the University of Cambridge University Information Services on behalf of the STFC DiRAC HPC Facility (\url{www.dirac.ac.uk}{www.dirac.ac.uk}). The DiRAC component of CSD3 at Cambridge was funded by BEIS, UKRI and STFC capital funding and STFC operations grants. DiRAC is part of the UKRI Digital Research Infrastructure. This work started as a collaboration during the WITGAF 2019 workshop in Cargese; the authors thank the organisers of the workshop. 
The authors thank the reviewers for suggestions that have improved the manuscript.

\end{document}


%
%


\title{Supporting Information for ``The cross-over from viscous to inertial lengthscales in rapidly-rotating convection"}
%
%

%
%



\authors{C. Guervilly\affil{1}, and E. Dormy\affil{2}}


\affiliation{1}{School of Mathematics, Statistics and Physics, Newcastle University, Newcastle Upon Tyne NE1 7RU, United Kingdom}
\affiliation{2}{Department of Mathematics \& its Applications, \'Ecole Normale Sup\'erieure, CNRS, PSL University, 75005 Paris, France}

%
%

%

\begin{article}

%
%

\noindent\textbf{Contents of this file}
\begin{enumerate}
\item Figures S1 to S2
\item Tables S1 to S2
\end{enumerate}


\clearpage

\centering
\includegraphics[width=\textwidth]{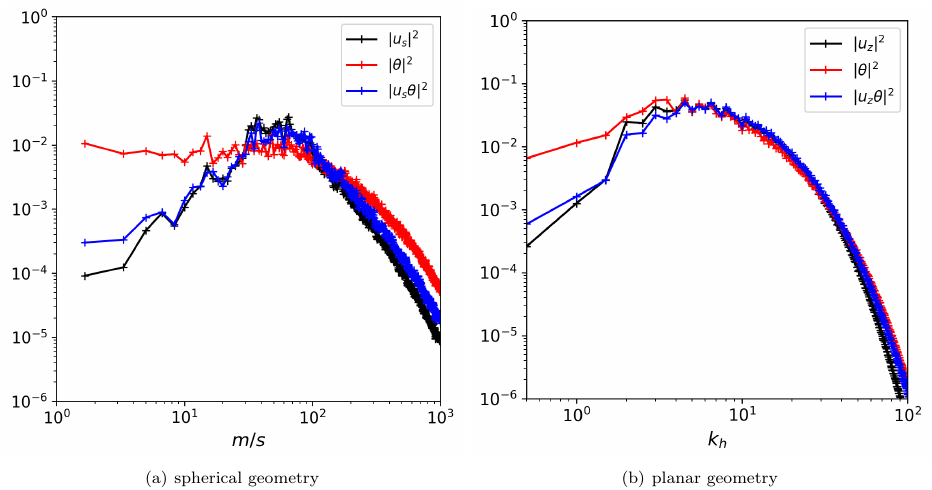}\\
{\bf Figure S1: } Power spectra of the kinetic energy of the radial/vertical velocity, temperature perturbation,
and convective flux as a function of (a) the azimuthal wavenumber $m/s$ in a QG spherical simulation and (b)
the horizontal wavenumber $k_h=(k_x^2+k_y^2)^{1/2}$.
The parameters for the QG simulation in (a) are $\Ek=10^{-8}$, $\Pran=1$,
$\Ra=4.45\times10^{11}$ (corresponding to $\Ro^{1/2}/\Ek^{1/3}=4.44$).
The parameters for the planar simulation in (b) are $\Ek=2\times 10^{-5}$, $\Pran=1$,
$\Ra=5\times10^8$ (corresponding to $\Ro^{1/2}/\Ek^{1/3}=5.57$) and $h=2d$.
Each spectrum is normalised by the sum of the plotted quantity over all the wavenumbers.
The spectra are calculated from a data snapshot at a given time and averaged
in the radial/vertical direction, excluding the boundary layers.
Accordingly, in (a), the distribution is plotted as a function of the  
horizontal wavenumber $m/s$.

\newpage
\centering
\includegraphics[width=0.8\textwidth]{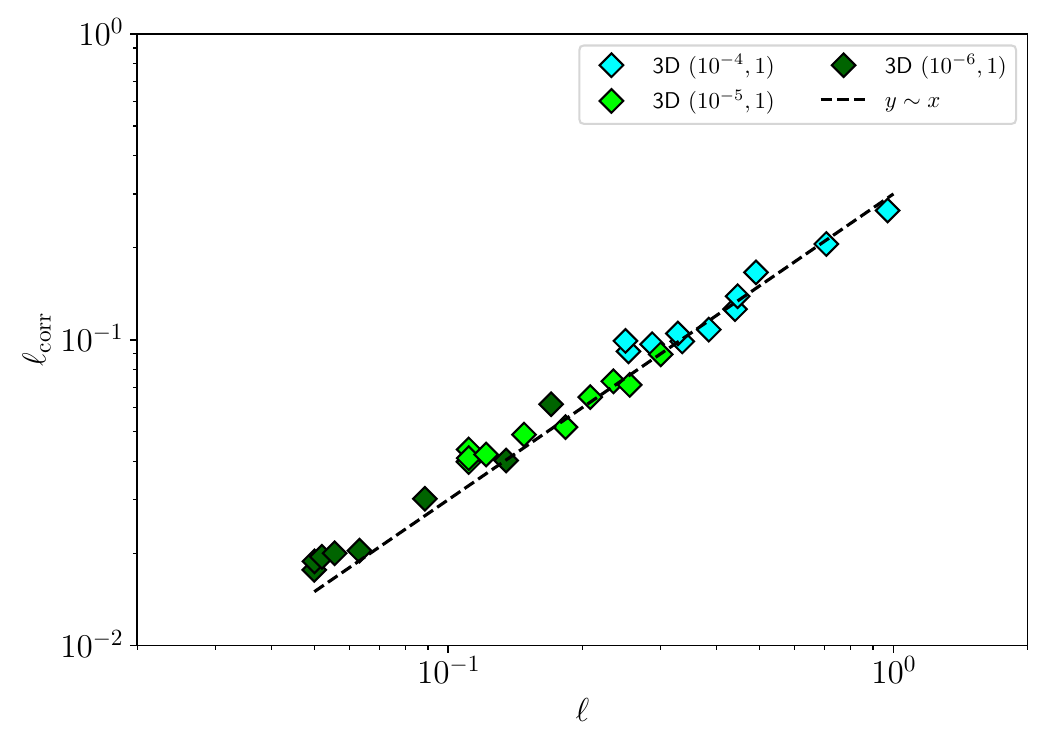}\\
{\bf Figure S2: } Comparison of the correlation lengthscale $\ell_{\rm corr}$ and the lengthscale calculated from the
spectral peak $\lu$ in planar simulations. 
The correlation lengthscale is calculated from the 2D auto-correlation functions $f(d_x,d_y,z)$ of the vertical velocity
in the $xy$ planes at all depth $z$ (excluding the boundary layers). 
The data is sorted into bins of horizontal length $(d_x^2+d_y^2)^{1/2}$ and
$\ell_{\rm corr}$ is computed as the width of the first zero of the auto-correlation function and averaged in depth. 
The calculation is made from a data snapshot at a given time, but 
the procedure was repeated at different times to check that the 
data are representative.


\newpage 
\centering
\begin{tabular}{c c c c c c c c c}
\hline
\hline
$\Ek$  & $\Pran$ & $\Ra$ & $\widetilde{\Ra}$ & $h/d$ & $n_x \times n_y \times n_z$ & $\Rey$ & $\lt/\lnu$ & $\lu$ \\
\hline
\hline
$2\times10^{-4}$  & $1$ & $1.6\times10^6$ & $18.7$ & $4$ & $256 \times 256 \times 64$ & $9.6$ & $0.74$ & $0.250$ \\
$2\times10^{-4}$  & $1$ & $1.8\times10^6$ & $21.1$ & $4$ & $256 \times 256 \times 64$ & $21.1$ & $1.11$ & $0.250$ \\
$2\times10^{-4}$  & $1$ & $2.2345\times10^6$ & $26.1$ & $4$ & $256 \times 256 \times 64$ & $41.4$ & $1.56$ & $0.254$ \\
$2\times10^{-4}$  & $1$ & $3\times10^6$ & $35.1$ & $4$ & $256 \times 256 \times 64$ & $71.5$ & $2.04$ & $0.287$ \\
$2\times10^{-4}$  & $1$ & $4\times10^6$ & $46.88$ & $4$ & $256 \times 256 \times 96$ & $103.6$ & $2.46$ & $0.293$ \\
$2\times10^{-4}$  & $1$ & $6\times10^6$ & $70.17$ & $4$ & $256 \times 256 \times 96$ & $153.2$ & $2.99$ & $0.335$ \\
$2\times10^{-4}$  & $1$ & $8\times10^6$ & $93.6$ & $4$ & $256 \times 256 \times 96$  & $191.9$ & $3.35$ & $0.328$ \\
$2\times10^{-4}$  & $1$ & $1.5\times10^7$ & $175.4$ & $4$ & $256 \times 256 \times 96$ & $287.4$ & $4.10$ & $0.385$ \\
$2\times10^{-4}$  & $1$ & $3\times10^7$ & $350.9$ & $4$ & $256 \times 256 \times 96$ & $430.8$ & $5.02$ & $0.440$ \\
$2\times10^{-4}$  & $1$ & $5\times10^7$ & $584.8$ & $4$ & $256 \times 256 \times 96$ & $571.0$ & $5.78$ & $0.446$ \\
$2\times10^{-4}$  & $1$ & $7.5\times10^7$ & $877.2$ & $4$ & $256 \times 256 \times 96$ & $708.9$ & $6.44$ & $0.491$ \\
$2\times10^{-4}$  & $1$ & $10^8$ & $1169.6$ & $4$ & $256 \times 256 \times 96$ & $828.1$ & $6.96$ & $0.550$ \\
$2\times10^{-4}$  & $1$ & $2\times10^8$ & $2339.2$ & $4$ & $324 \times 324 \times 128$ & $1182.1$ & $8.31$ & $0.706$ \\
$2\times10^{-4}$  & $1$ & $4\times10^8$ & $4678.4$ & $4$ & $324 \times 324 \times 128$ & $1758.6$ & $10.14$ & $0.970$ \\
\hline
$2\times10^{-5}$  & $1$ & $3.5\times10^7$ & $19.0$ & $1$ & $256 \times 256 \times 128$ & $11.5$ & $0.56$ & $0.111$ \\
$2\times10^{-5}$  & $1$ & $4\times10^7$ & $21.7$ & $1$ & $256 \times 256 \times 128$ & $43.8$ & $1.09$ & $0.111$ \\
$2\times10^{-5}$  & $1$ & $4\times10^7$ & $21.7$ & $2$ & $512 \times 512 \times 128$ & $43.3$ & $1.08$ & $0.111$ \\
$2\times10^{-5}$  & $1$ & $4.17\times10^7$ & $22.6$ & $1$ & $256 \times 256 \times 128$ & $52.3$ & $1.19$ & $0.111$ \\
$2\times10^{-5}$  & $1$ & $5\times10^7$ & $27.1$ & $1$ & $256 \times 256 \times 128$ & $94.0$ & $1.60$ & $0.111$ \\
$2\times10^{-5}$  & $1$ & $6.96\times10^7$ & $37.8$ & $1$ & $256 \times 256 \times 128$ & $196.4$ & $2.31$ & $0.122$ \\
$2\times10^{-5}$  & $1$ & $1.39\times10^8$ & $75.5$ & $1$ & $256 \times 256 \times 192$ & $480.2$ & $3.61$ & $0.148$ \\
$2\times10^{-5}$  & $1$ & $2.32\times10^8$ & $125.9$ & $1$ & $256 \times 256 \times 192$ & $684.2$ & $4.31$ & $0.183$ \\
$2\times10^{-5}$  & $1$ & $3.25\times10^8$ & $176.4$ & $1$ & $256 \times 256 \times 256$ & $853.8$ & $4.81$ & $0.208$ \\
$2\times10^{-5}$  & $1$ & $4\times10^8$ & $217.1$ & $1$ & $256 \times 256 \times 256$ & $987.3$ & $5.18$ & $0.235$ \\
$2\times10^{-5}$  & $1$ & $5\times10^8$ & $271.4$ & $1$ & $256 \times 256 \times 256$ & $1148.9$ & $5.58$ & $0.256$ \\
$2\times10^{-5}$  & $1$ & $5\times10^8$ & $271.4$ & $2$ & $512 \times 512 \times 256$ & $1143.0$ & $5.57$ & $0.219$ \\
$2\times10^{-5}$  & $1$ & $8\times10^8$ & $434.3$ & $1$ & $256 \times 256 \times 256$ & $1557.4$ & $6.50$ & $0.300$ \\
$2\times10^{-5}$  & $1$ & $10^9$ & $542.9$ & $2$ & $512 \times 512 \times 256$ & $1779.9$ & $6.95$ & $0.350$ \\
$2\times10^{-5}$  & $1$ & $2\times 10^9$ & $1085.7$ & $2$ & $512 \times 512 \times 256$ & $2650.5$ & $8.48$ & $0.454$ \\
\hline
$2\times10^{-6}$  & $1$ & $8.5\times10^8$ & $21.4$ & $0.5$ & $256 \times 256 \times 192$ & $67.1$ & $0.92$ & $0.0500$ \\
$2\times10^{-6}$  & $1$ & $9\times10^8$ & $22.7$ & $0.5$ & $256 \times 256 \times 192$ & $89.6$ & $1.06$ & $0.0500$ \\
$2\times10^{-6}$  & $1$ & $10^9$ & $25.2$ & $0.5$ & $256 \times 256 \times 192$ & $144.2$ & $1.35$ & $0.0520$ \\
$2\times10^{-6}$  & $1$ & $1.5\times10^9$ & $37.8$ & $0.5$ & $256 \times 256 \times 192$ & $395.0$ & $2.23$ & $0.0555$ \\
$2\times10^{-6}$  & $1$ & $2\times10^9$ & $50.4$ & $0.5$ & $256 \times 256 \times 192$ & $705.1$ & $2.98$ & $0.0631$ \\
$2\times10^{-6}$  & $1$ & $4\times10^9$ & $100.8$ & $0.5$ & $256 \times 256 \times 192$ & $1416.0$ & $4.22$ & $0.0886$ \\
$2\times10^{-6}$  & $1$ & $6\times10^9$ & $151.2$ & $0.5$ & $256 \times 256 \times 256$ & $1948.5$ & $4.95$ & $0.135$ \\
$2\times10^{-6}$  & $1$ & $8\times10^9$ & $201.6$ & $0.5$ & $256 \times 256 \times 384$ & $2488.8$ & $5.60$ & $0.153$ \\
$2\times10^{-6}$  & $1$ & $10^{10}$ & $252.0$ & $0.5$ & $256 \times 256 \times 384$ & $3082.6$ & $6.23$ & $0.170$ \\
\hline
\hline
\end{tabular}
\vspace{0.4cm}
{\bf Table S1: } Parameters and output quantities for the planar simulations.
The reduced Rayleigh number is defined as $\widetilde{\Ra}=\Ra\Ek^{4/3}$.
$h/d$ is the aspect ratio of the computational domain.
The numerical resolution is given as $n_x$ and $n_y$ Fourier modes for the $x$ and $y$ directions
and $n_z$ modes for Chebyshev polynomials in the $z$ direction.
The Reynolds number $\Rey$ is based on the value of vertical velocity averaged in volume and time. 
The ratio of the inertial to viscous scaling is $\lt/\lnu=\Ro^{1/2}/\Ek^{1/3}$, where the Rossby number is $\Ro=\Rey\Ek$.

\newpage 
\centering
\begin{tabular}{c c c c c c c c}
\hline
\hline
$\Ek$  & $\Pran$ & $\Ra$ & $\widetilde{\Ra}$ & $m_{\rm max} \times n_r $ & $\Rey$ & $\lt/\lnu$ & $\lu$ \\
\hline
\hline
$10^{-8}$  & $1$ & $2.225\times10^{10}$ & $0.48$ &  $600 \times 4000$ & $58.9$ & $0.36$ & $0.0106$ \\
$10^{-8}$  & $1$ & $2.67\times10^{10}$ & $0.58$ &  $600 \times 4000$ & $85.0$ & $0.43$ & $0.0123$ \\
$10^{-8}$  & $1$ & $3.56\times10^{10}$ & $0.77$ &  $600 \times 4000$ & $264.1$ & $0.75$ & $0.0118$ \\
$10^{-8}$  & $1$ & $4.45\times10^{10}$ & $0.96$ &  $600 \times 4000$ & $409.9$ & $0.94$ & $0.0122$ \\
$10^{-8}$  & $1$ & $6.675\times10^{10}$ & $1.4$ &  $600 \times 4000$ & $872.5$ & $1.37$ & $0.0144$ \\
$10^{-8}$  & $1$ & $8.9\times10^{10}$ & $1.9$ &  $600 \times 4000$ & $1314.1$ & $1.68$ & $0.0182$ \\
$10^{-8}$  & $1$ & $1.78\times10^{11}$ & $3.8$ &  $600 \times 4000$ & $2863.9$ & $2.48$ & $0.0313$ \\
$10^{-8}$  & $1$ & $2.67\times10^{11}$ & $5.8$ &  $600 \times 4000$ & $4979.4$ & $3.28$ & $0.0402$ \\
$10^{-8}$  & $1$ & $3.56\times10^{11}$ & $7.7$ &  $800 \times 4000$ & $7824.9$ & $4.11$ & $0.0475$ \\
$10^{-8}$  & $1$ & $4.45\times10^{11}$ & $9.6$ &  $800 \times 4000$ & $9151.4$ & $4.44$ & $0.0515$ \\
\hline
$10^{-8}$  & $0.01$ & $1.78\times10^{9}$ & $0.038$ &  $200 \times 2000$ & $5372.1$ & $3.40$ & $0.0479$ \\
$10^{-8}$  & $0.01$ & $3.56\times10^{9}$ & $0.077$ &  $300 \times 2000$ & $13161.3$ & $5.32$ & $0.0579$ \\
$10^{-8}$  & $0.01$ & $6.675\times10^{9}$ & $0.144$ &  $300 \times 2000$ & $23267.5$ & $7.08$ & $0.0714$ \\
$10^{-8}$  & $0.01$ & $1.1125\times10^{10}$ & $0.240$ &  $384 \times 2500$ & $36404.2$ & $8.86$ & $0.0818$ \\
\hline
\hline
\end{tabular}
\vspace{0.4cm}
{\bf Table S2: } Parameters and output quantities for the QG simulations with differential heating.
The numerical resolution is given as $m_{\rm max}$ Fourier modes for the azimuthal direction
and $n_r$ radial points in the radial direction.
The Reynolds number $\Rey$ is based on the value of radial velocity averaged in volume and time. 

\end{article}
\clearpage